\documentclass[twocolumn,preprintnumbers,prb,amsmath,amssymb,superscriptaddress]{revtex4-1}
\usepackage{graphicx}
\usepackage{dcolumn}
\usepackage{bm}
\usepackage{titlesec}

\begin{document}

\title{Flux growth in a horizontal configuration: an analogue to vapor transport growth}

\author{J.-Q. Yan}
\email{yanj@ornl.gov}
\affiliation{Materials Science and Technology Division, Oak Ridge National Laboratory, Oak Ridge, Tennessee 37831, USA}

\author{B. C. Sales}
\affiliation{Materials Science and Technology Division, Oak Ridge National Laboratory, Oak Ridge, Tennessee 37831, USA}

\author{M. A. Susner}
\affiliation{Materials Science and Technology Division, Oak Ridge National Laboratory, Oak Ridge, Tennessee 37831, USA}
\affiliation{Aerospace Systems Directorate, Air Force Research Laboratory 1950 Fifth St., Building 18, Wright-Patterson Air Force Base, OH 45433 USA}
\affiliation{UES, Inc. 4401 Dayton Xenia Rd. Beavercreek, OH 45432 USA}

\author{M. A. McGuire}
\affiliation{Materials Science and Technology Division, Oak Ridge National Laboratory, Oak Ridge, Tennessee 37831, USA}
\date{\today}

\begin{abstract}
Flux growth of single crystals is normally performed in a vertical configuration with an upright refractory container holding the flux melt. At high temperatures,  flux dissolves the charge forming a homogeneous solution before nucleation and growth of crystals take place under proper supersaturation generated by cooling or evaporating the flux. In this work, we report flux growth in a horizontal configuration with a temperature gradient along the horizontal axis: a liquid transport growth analogous to the vapor transport technique. In a typical liquid transport growth, the charge is kept at the hot end of the refractory container and the flux melt dissolves the charge and transfers it to the cold end. Once the concentration of charge is above the solubility limit at the cold end, the thermodynamically stable phase nucleates and grows. Compared to the vertical flux growth, the liquid transport growth can provide a large quantity of crystals in a single growth since the charge/flux ratio is not limited by the solubility limit at the growth temperature. This technique is complementary to the vertical flux growth and can be considered when a large amount of crystals are needed but the yield from the conventional vertical flux growth is limited. We applied this technique to the growth of IrSb$_3$, Mo$_3$Sb$_7$, MnBi from self flux, and the growth of FeSe, CrTe$_3$, NiPSe$_3$, FePSe$_3$, and InCuP$_2$S$_6$ from a halide flux.

\end{abstract}

\maketitle
\section{Introduction}
Flux growth is a compelling method for the discovery of new materials and for the growth of crystals for scientific studies as well as technological applications.\cite{canfield2001high,kanatzidis2005metal,yan2015flux,phelan2011adventures,wanklyn1972flux,bugaris2012materials}  In a typical flux growth, the mixture of target compound and a flux with a low melting point is heated to high temperatures where the flux completely dissolves other starting materials, thereby forming a homogeneous solution. Then, the temperature is slowly lowered to generate supersaturation necessary for the nucleation and growth of crystals. Crystal growth can also take place under a supersaturation generated by evaporating the flux, which takes advantage of the large vapor pressures of some fluxes. A refractory container, such as the Al$_2$O$_3$-based Canfield crucible set,\cite{canfield2016use} is needed to hold the high temperature melt. The refractory crucible is normally placed in an upright position to keep the melt inside and no temperature gradient, either vertical or radial, is purposely generated. In the following text, we refer to this type of flux growth in a vertical configuration as "conventional vertical flux growth".

In this  paper, we report on flux growth in a horizontal configuration with a temperature gradient purposely applied along the horizontal axis. This technique is similar to vapor transport growth but here the flux melt behaving as the transport agent. We thus name it liquid transport growth. Quite different from that in the conventional vertical flux growth, the charge/flux ratio is not limited by the solubility of charge in the flux at given temperatures. Thus the liquid transport growth can produce a large quantity of single crystals in a single growth which is otherwise impossible for the conventional vertical flux growth. This technique is complementary to the conventional vertical flux growth and can be employed when a large amount of crystals are needed for some special measurements but the yield from the conventional vertical flux growth is limited. We show the growth details using this technique of IrSb$_3$, Mo$_3$Sb$_7$, and MnBi out of self flux, FeSe, CrTe$_3$, NiPSe$_3$, FePSe$_3$, and InCuP$_2$S$_6$ out of an AlCl$_3$-KCl eutectic flux. In addition, the reaction between transition metal oxides and the AlCl$_3$-KCl eutectic flux can provide clean small RuCl$_3$ crystals and OsCl$_4$ starting materials for further crystal growth of trihalides. The successful growth of these proof of principle compounds suggests that the liquid transport growth works well for a large variety of compounds. Temperature and the temperature gradient along the horizontal axis are critical growth parameters which determine the composition of the crystallizing phase which is thermodynamically stable at given temperatures and also affect the growth kinetics.

\section{Liquid transport vs vapor transport}

\begin{figure} \centering \includegraphics [width = 0.47\textwidth] {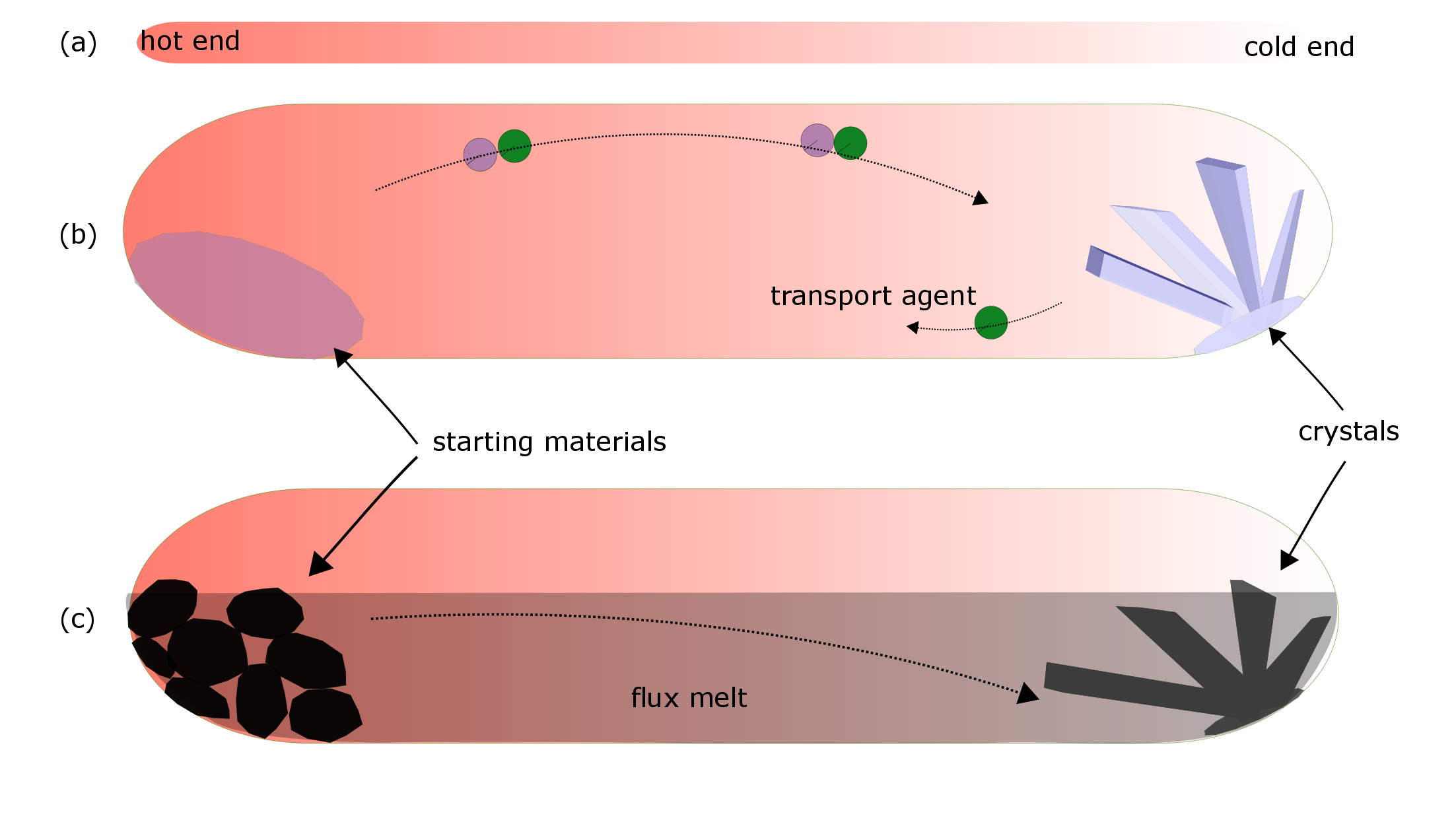}
\caption{(color online) (a) Temperature profile along the horizontal axis. (b)Schematic picture of vapor transport growth. (c) Schematic picture of liquid transport growth. }
\label{Transport}
\end{figure}

Figure\,\ref{Transport} illustrates the principles of vapor transport growth and liquid transport growth. A horizontal temperature gradient is applied for both techniques. In a typical vapor transport growth, a volatile substance (the so-called transport agent) is sealed in, for example, a quartz tube together with the raw materials. The sealed ampoule is then placed inside of a tube furnace with a well-defined temperature profile (See Fig.\,\ref{Transport}(a)). At high temperatures, the volatile transport agent reacts with the nonvolatile starting materials into some intermediate molecular species that can diffuse to the cold end. Once the partial pressure of the intermediate molecular species is high enough, crystallization takes place at the cold end and the transport agent is released. As the transport agent is not consumed in the growth, only a small amount of transport agent is needed. For example, an empirical amount of 3-5mg per cm$^3$ is normally used. For detailed description of this technique, see Reference [\cite{binnewies2013chemical,schmidt2013chemical}].

Liquid transport growth is similar to the vapor transport technique as illustrated in Fig.\,\ref{Transport}(c). The charge is kept at the hot end of the growth ampoule. In addition to the charge, a large amount of flux is added to partially fill the growth ampoule. At high temperatures, the flux dissolves the charge at the hot end and transfers the dissolved charge to the cold end driven by the composition gradient. At the cold end, once the concentration of the dissolved charge goes beyond the solubility limit, the thermodynamically stable phase starts to precipitate. As in the conventional vertical flux growth, the flux can be the constituent of the desired compound, i.e., the so-called self flux. In this case, the flux will be consumed during the crystal growth. The flux can also have compositions quite different from the starting materials and the desired crystal. In this case, the flux is not consumed during crystal growth but works as a liquid transport agent.

We first present the growth details of some proof of principle compounds in Section III and then discuss the growth mechanism of the liquid transport growth in Section IV.

\section{Liquid transport growth examples}

\subsection{liquid transport growth from self flux}

\subsubsection{IrSb$_3$}
\begin{figure} \centering \includegraphics [width = 0.47\textwidth] {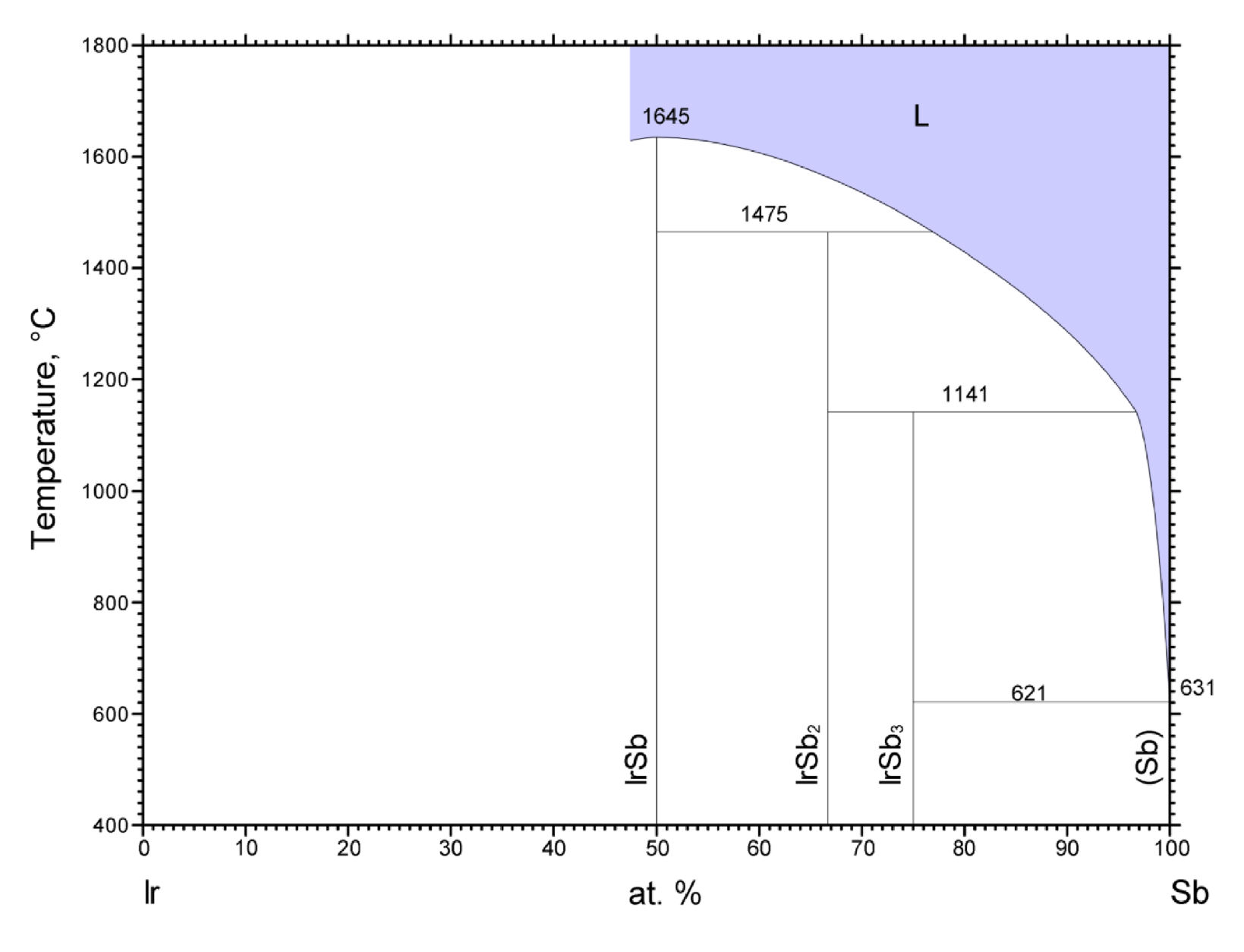}
\caption{(color online) Phase diagram of Ir-Sb from Reference[\citenum{IrSbPhaseDiagram}]. Other versions of phase diagram in the same database suggest the peritectic plateau extends to $\sim$85\% Sb. Our flux growths using the vertical configuration support the version shown here.}
\label{IrSbPD}
\end{figure}

IrSb$_3$ crystallizes in the Skutterudite structure (space group Im$\overline{3}$), which can host excellent thermoelectric properties upon filling the voids in the structure.\cite{sales1996filled} Recently, isostructural RhSb$_3$ was proposed to be a new zero-gap three-dimensional Dirac semimetal.\cite{wang2017dirac} This suggests that the unfilled skutterudite compounds provide a new materials playground for the study of topological Dirac semimetals. Figure\,\ref{IrSbPD} shows the Ir-Sb binary phase diagram.\cite{IrSbPhaseDiagram} The peritectic plateau extends to a composition of $\sim$97\% Sb, leaving a narrow composition range for the growth of IrSb$_3$ out of Sb flux and the liquidus line has a large composition dependence in this range. Nevertheless, we have managed to grow IrSb$_3$ single crystals out of Sb flux using the vertical configuration. However, less than 3\,\% Ir is allowed in the melt according to the phase diagram, which limits the amount of crystals one may obtain from a single growth. We thus tried flux growth in a horizontal configuration.

\begin{figure} \centering \includegraphics [width = 0.47\textwidth] {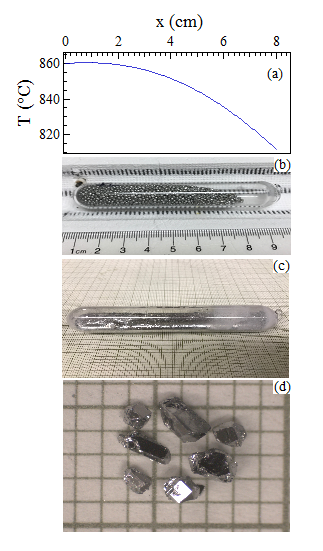}
\caption{(color online) Liquid transport growth of IrSb$_3$. (a) The temperature profile along the horizontal axis where the thermocouple is located at 0\,cm.  (b) The growth ampoule inside of the single zone tube furnace before heating up; the Ir powder is located at the hot end (x=0). (c) The ampoule after growth on a mm grid; single crystals were found in the 2-3\,cm range from the cold end. (d) IrSb$_3$ single crystals on a mm grid after removing the extra Sb flux as described in text.}
\label{IrSb}
\end{figure}

The starting materials are Ir powder (Alfa Aesar, -22 mesh, 99.99\%) and Sb shot (Alfa Aesar, 1-3\,mm, 99.9999\%). In a typical growth, about 1.5\,g Ir powder and 50\,g Sb shot were loaded into a quartz tube of 16\,mm inner diameter and 1.5\,mm wall thickness. The quartz tube was sealed under vacuum (see Fig.\,\ref{IrSb}(b)). The growth ampoule was then put in a single zone tube furnace with the Ir powder located at the hot end. It is worth mentioning that mixing of Ir and Sb starting materials should be minimized. Figure\,\ref{IrSb}(a) shows the temperature profile along the tube. After keeping the hot end (x=0) at 860$^\circ$C for two weeks, the furnace was turned off. Figure\,\ref{IrSb}(c) shows the ampoule after growth. IrSb$_3$ crystals were found within about $\sim$3\,cm of the cold end of the tube. To isolate the crystals from the extra Sb flux, we sealed the part cut from the cold end in a Canfield crucible set under vacuum and heated it to 650$^\circ$C. After dwelling at this temperature for two hours, the extra Sb flux was removed by centrifuging. Figure\,\ref{IrSb}(d) shows some millimeter-sized IrSb$_3$ crystals left in the growth crucible after decanting. The crystals are of similar size as those obtained from the conventional vertical flux technique. In such a growth, we obtain about 2.8\,g of IrSb$_3$ crystals. This is much more than what one can grow using the conventional vertical flux growth. In a typical vertical flux growth using a 2\,ml Canfield crucible set, about 0.4\,g of  IrSb$_3$ crystals can be obtained starting with a 4\,g batch containing 0.18\,g of Ir as the charge. The liquid transport grown IrSb$_3$  crystals exhibit the same transport properties as those grown in the conventional vertical configuration. The detailed physical properties of IrSb$_3$ crystals will be reported elsewhere as it is not the focus of the present work.

\begin{figure} \centering \includegraphics [width = 0.47\textwidth] {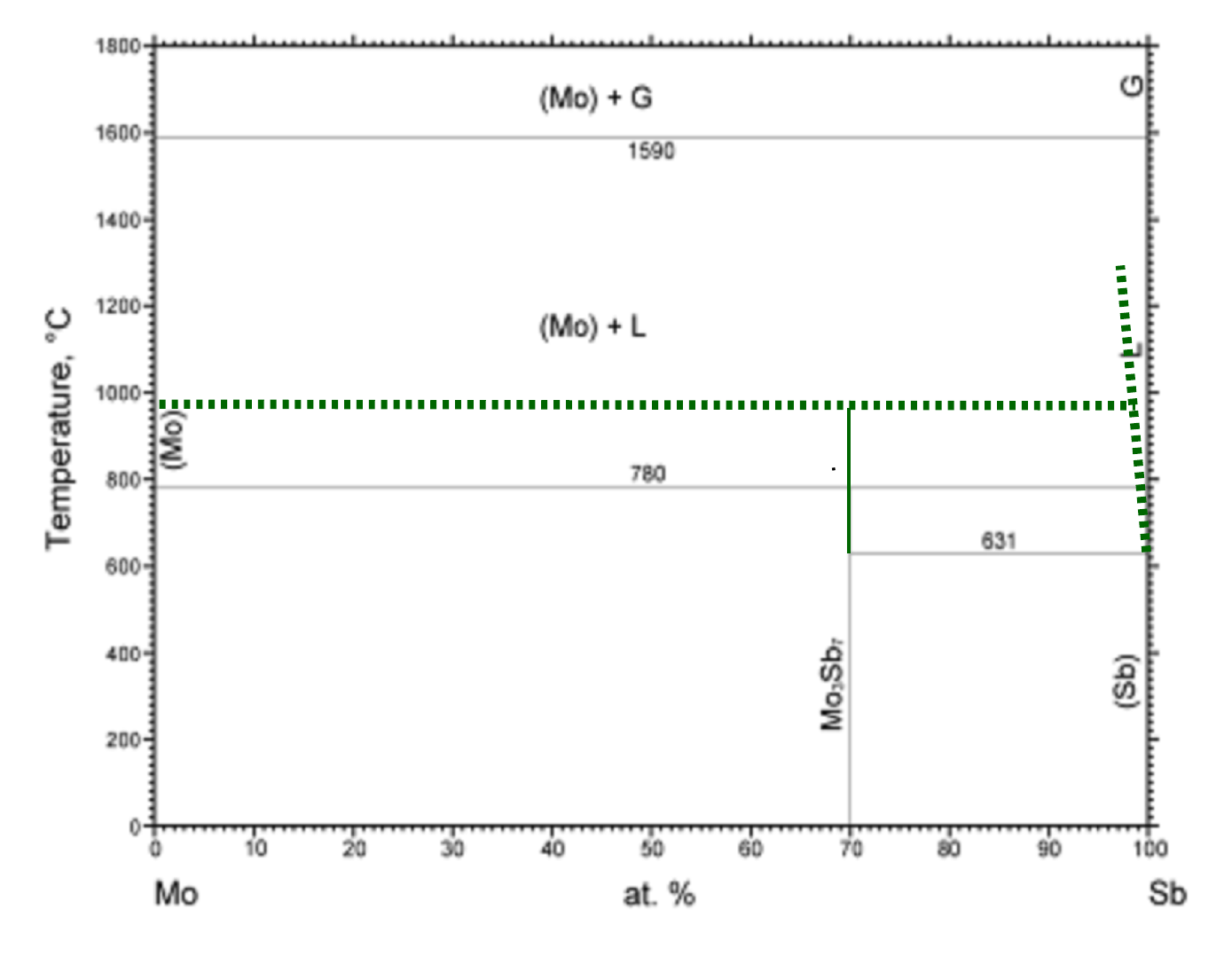}
\caption{(color online) Phase diagram of Mo-Sb from Reference[\citenum{IrSbPhaseDiagram}]. The dashed curve and line are added following our flux growths and Differential Scanning Calorimetry measurements.}
\label{MoSb}
\end{figure}

\begin{figure} \centering \includegraphics [width = 0.47\textwidth] {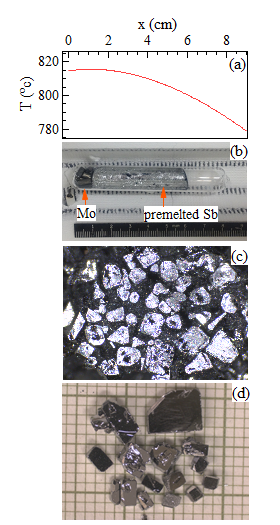}
\caption{(color online) Liquid transport growth of Mo$_3$Sb$_7$.(a) The temperature gradient along the horizontal axis. (b) The ampoule before growth with Mo and premelted Sb sealed in a quartz tube. (c) Single crystals buried in Sb flux. (d) Mo$_3$Sb$_7$ crystals on a millimeter grid after mechanically removing the Sb flux using a surgical blade.}
\label{Mo3Sb7}
\end{figure}

\subsubsection{Mo$_3$Sb$_7$}
Despite a low superconducting transition temperature of 2.30 K, the Zintl compound Mo$_3$Sb$_7$ attracted much attention due to possible interplay between superconductivity, structural instability, and magnetism. With appropriate doping, Mo$_3$Sb$_7$ also shows good thermoelectric performance. The first Mo$_3$Sb$_7$ single crystal large enough for physical property measurements was grown by melting Sb in a thick Mo tube by Bulowski et al.\cite{bukowski2002single} This work motivated us to grow millimeter sized Mo$_3$Sb$_7$ single crystals out of Sb flux.\cite{yan2013flux} In a typical flux growth using the conventional vertical configuration, Mo slug or reduced powder and Sb shot were mixed in the molar ratio of Mo$:$Sb = 1$:$49 and placed in a 2\,ml Al$_2$O$_3$ crucible. A catch crucible of the same size containing quartz wool was mounted on top of the growth crucible and both were sealed in a silica ampoule under approximately 1/3 atmosphere of argon gas. The sealed ampoule was heated to 1000$^\circ$C in 5 hours and homogenized for 12 hours before cooling to 700$^\circ$C over 100 hours in a programmable box furnace. At 700$^\circ$C, the Sb flux was decanted from the Mo$_3$Sb$_7$ crystals. A similar procedure has been successfully employed to grow doped single crystals.\cite{yan2015fragile}

The above self flux technique can produce Mo$_3$Sb$_7$ single crystals with the largest dimension around 3\,mm. However, the small fraction of Mo in the starting composition and the low yield of sizeable pieces make it very time consuming and expensive to grow enough pieces for inelastic neutron scattering measurements for the study of lattice dynamics and possible magnetic resonance.

Figure\,\ref{MoSb} shows the Mo-Sb binary phase diagram.\cite{IrSbPhaseDiagram} The successful growth of Mo$_3$Sb$_7$ crystals out of Sb flux suggests the reported phase diagram misses the liquidus line. Our Differential Scanning Calorimetry measurements suggest the peritectic temperature is above 950$^\circ$C. We necessarily modified the phase diagram by adding a dashed line denoting the peritectic reaction and a dashed curve as the liquidus line.  Mo$_3$Sb$_7$ is the only compound in the Mo-Sb binary system. Following the idea of liquid transport technique described in Section II, we would expect Sb flux reacts with Mo  at the hot end and deposites Mo$_3$Sb$_7$ at the cold end as long as the cold end is kept below the peritectic temperature.

Figure\,\ref{Mo3Sb7}(a) shows the temperature profile used in the liquid transport growth of Mo$_3$Sb$_7$. The temperature profile along the horizontal axis was measured after the furnace dwells at the setup temperature for at least half an hour. First, about 6.5\,g of Mo slugs (Alfa Aesar, 99.95\%) and 50\,g of Sb shot (Alfa Aesar, 1-3\,mm, 99.9999\%) were loaded into a quartz tube of 12\,cm length, 14\,mm inner diameter and 1.5\,mm wall thickness. Similar to the aforementioned growth of IrSb$_3$, the mixing of Mo and Sb starting materials should be avoided. Using Sb ingots premelted in tubes of same inner diameter can help control the tube filling and avoid mixing of Mo slug and Sb shot. The quartz tube was then sealed under vacuum and the sealed ampoule was put in a single zone tube furnace with Mo located at the hot end (Fig.\,\ref{Mo3Sb7}(b)). The furnace was set up at 815$^\circ$C and kept at this temperature for a week before furnace cooling to room temperature.

Mo$_3$Sb$_7$ crystals were found in the whole tube and buried in Sb flux. Figure\,\ref{Mo3Sb7}(c) shows the picture of one small portion of ingot with Mo$_3$Sb$_7$ crystals buried inside of Sb flux. Sb flux around crystals can be mechanically removed using a surgical blade. Over 15\,g Mo$_3$Sb$_7$ crystals with the largest dimension up to 5\,mm can be isolated from the Sb flux as shown in Figure\,\ref{Mo3Sb7}(d). The largest crystals are normally found at the cold end and the average size of the crystals decreases from the cold end to the hot end. Crystals at the cold end are of comparable size to those grown in the vertical configuration.  Crystals from different positions were purposely picked up and characterized by measuring x-ray powder diffraction, magnetic susceptibility, and electrical resistivity. No sample variation was observed. In addition, single crystals grown in this technique have the same physical properties as those grown out of Sb flux using traditional vertical configuration.\cite{yan2013flux,yan2015fragile}

Compared to the conventional vertical flux growth, the liquid transport growth produces a much larger quantity of Mo$_3$Sb$_7$ crystals of comparable dimension in a single growth. In a typical vertical flux growth using a 2\,ml Canfield crucible set, about 0.2\,g of  Mo$_3$Sb$_7$ crystals can be obtained starting with a 4\,g batch with 0.06\,g of Mo as the charge.

\begin{figure} \centering \includegraphics [width = 0.47\textwidth] {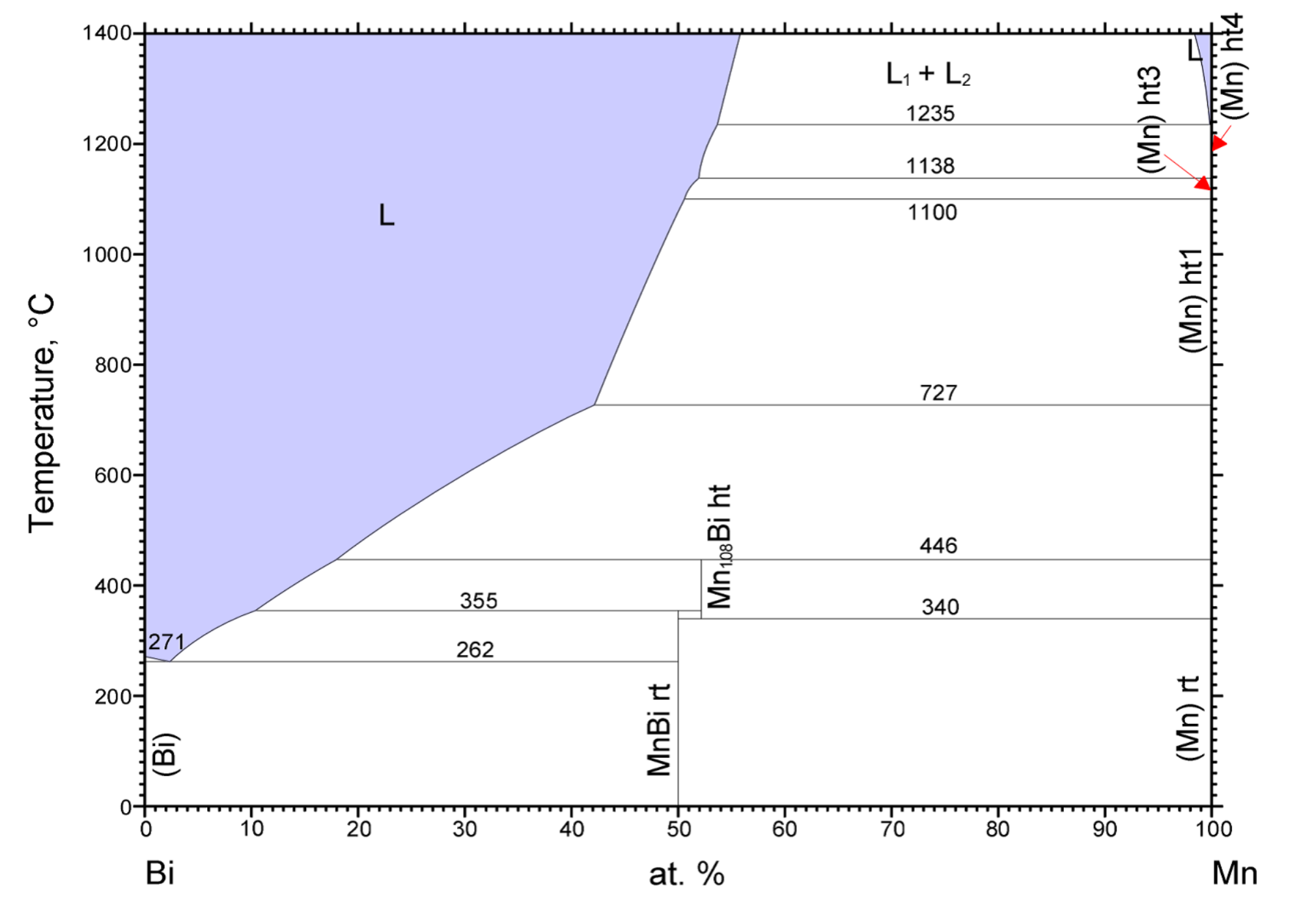}
\caption{(color online) Phase diagram of Mn-Bi from Reference[\citenum{IrSbPhaseDiagram}].}
\label{MnBiPD}
\end{figure}

\begin{figure} \centering \includegraphics [width = 0.47\textwidth] {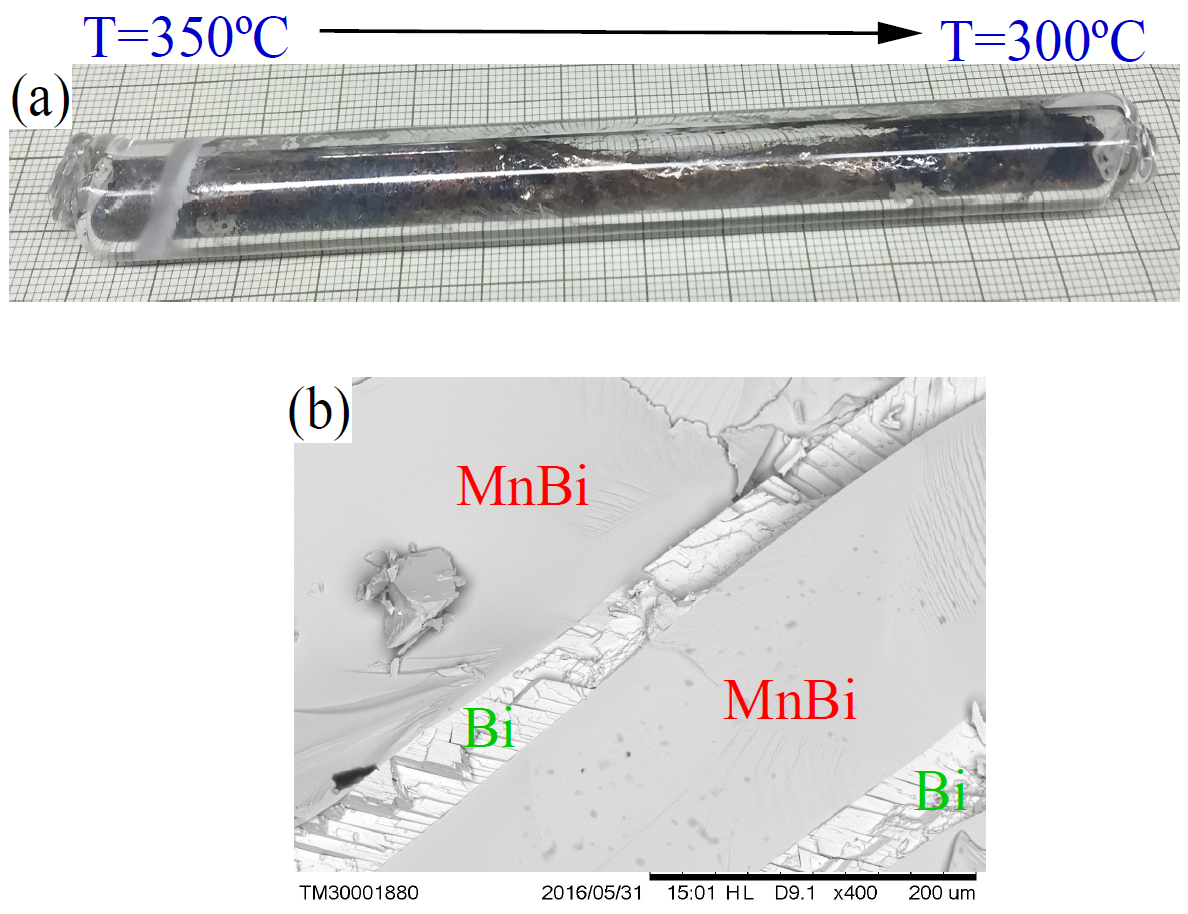}
\caption{(color online) Liquid transport growth of Mn-Bi. (a) ampoule after growth. The ampoule was tilted at about 300$^o$C when Bi is still in liquid state. MnBi was found in about 2-3\,cm range from the cold end. (b) The SEM picture of the cross-section of the precipitation at the cold end.}
\label{MnBi}
\end{figure}

\subsubsection{MnBi}
MnBi shows unusual magnetic and magneto-optical properties that has attracted interests for decades. There is recent revived interest in this compound because of its potential as a rare-earth-free permanent magnet at high temperatures and as a  hard phase in nanocomposite magnets.\cite{williams2016extended} Figure\,\ref{MnBiPD} shows the Mn-Bi binary phase diagram. MnBi melts incongruently at 355$^\circ$C and can be precipitated out of Bi$_{1-x}$Mn$_x$ (3$\leq$x$\leq$10) melt in the temperature range 262$\sim$355$^\circ$C. Compared to IrSb$_3$ and Mo$_3$Sb$_7$, the composition range appropriate for flux growth is wider but with a narrower temperature range. One more important feature that distinguishes MnBi from IrSb$_3$ and Mo$_3$Sb$_7$ is that the composition of MnBi (50\% Mn) is quite different from that of the coexisting liquid (3-10\% Mn). This limits the yield of conventional vertical flux growth following the lever rule. We thus tried the flux growth  using the horizontal configuration.

The starting materials are Bi shot (Alfa Aesar, spherical, 1-5\,mm, 99.999\%) and Mn pieces (Alfa Aesar, 99.95\%). Mn was arc melted before using to remove the surface oxide layer. In a typical growth, about 1g of Mn was first loaded into a quartz tube of 9\,mm inner diameter and 1.5\,mm wall thickness. Then about 10\,cm of the tube was filled with Bi shot. The ampoule was then sealed under vacuum and put into a single zone tube furnace with the hot end at 350$^\circ$C and the cold end at 300$^\circ$C. After one week, the ampoule was taken out of the furnace and cooled to room temperature.  Figure\,\ref{MnBi}(a) shows the ampoule after the growth. MnBi precipitation was found within about 3\,cm of the cold end. X-ray powder diffraction found both MnBi and Bi at the cold end. MnBi crystals are much smaller than those grown using the conventional vertical configuration. As the crystals are small, we thus looked at the fracture surface of the as-grown ingot at the cold end using scanning electron microscope (SEM). Figure \,\ref{MnBi}(b) shows the SEM picture in which submillimeter-sized MnBi crystals are separated by Bi flux. As discussed Section IV, the growth of MnBi crystal is limited by the viscosity of Bi melt at such low growth temperatures.

MnBi is not the only compound in the Mn-Bi binary system and MnBi is stable only in the temperature range 262$\sim$355$^\circ$C. Thus the growth temperature should be carefully controlled to avoid the precipitation of other phases. For example, above 355$^\circ$C, Mn$_{1.06}$Bi will form according to the phase diagram shown in Fig.\,\ref{MnBiPD}. We will further discuss the importance of temperature in section IV.

\subsection{Liquid transport growth from a halide flux}
\subsubsection{FeSe superconductor}
\begin{figure} \centering \includegraphics [width = 0.47\textwidth] {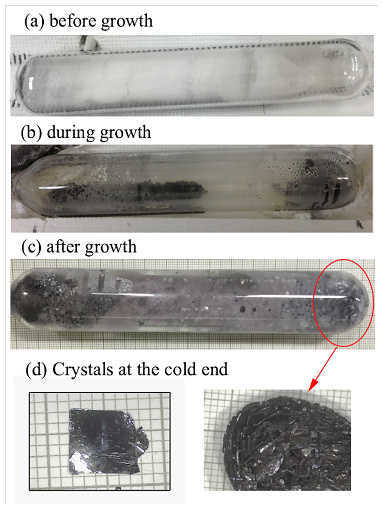}
\caption{(color online) Crystal growth of FeSe out of a KCl-AlCl$_3$ eutectic flux.(a) the ampoule before growth. (b) the ampoule during growth with some crystals at the cold end inside of the clear melt. (c) the ampoule after growth. (d) crystals at the cold end. A cluster of plate-like crystals were normally found at the cold end. The left panel shows a single piece of FeSe crystal on a mm grid.}
\label{FeSe}
\end{figure}

Among all Fe-based superconductors, FeSe has the simplest chemical formula and structure. However, crystal growth has been challenging because the superconducting tetragonal phase is stable only below $\sim$450$^\circ$C in a narrow compositional range and the superconductivity is extremely sensitive to the nonstoichiometry.\cite{mcqueen2009extreme} A KCl-AlCl$_3$ eutectic flux, which melts around 120$^\circ$C, was found to be ideal for the growth of high quality FeSe single crystals of millimeter size.\cite{chareev2013single} The growth was performed in a tube furnace with the sealed quartz ampoule sitting horizontally in a temperature gradient. Our growth of FeSe crystals follows the procedures reported by Chareev et al.\cite{chareev2013single} but starting with pre-reacted FeSe. The starting materials are iron powder (Alfa Aesar, -22 mesh, 99.998\%) and selenium shot (Alfa Aesar, amorphous, 2-6\,mm, 99.999\%). The starting materials in the molar ratio of Fe:Se=1.063:1 were sealed in a quartz tube under vacuum and heated in a box furnace to 250$^\circ$C in 10 hours, kept at this temperature for 12 hours, then heated to 750$^\circ$C in 24 hours and dwelled for 16hours, and finally heated to 1100$^\circ$C in 20 hours and homogenized at this temperature for 20 hours before furnace cooling to room temperature. The premelted FeSe was ground into coarse powder and loaded into a quartz tube with an inner diameter of 19\,mm and thickness of 1.5\,mm. Then AlCl$_3$ and KCl in a ratio of AlCl$_3$:KCl=2:1 were introduced into the ampoule. The step is done inside of a dry glove box filled with He as AlCl$_3$ is rather hygroscopic. The premelted FeSe is kept at one end of the tube and not mixed with the halide flux. The ampoule was then sealed under vacuum and Fig.\,\ref{FeSe}(a) shows a picture of such an ampoule before growth. The ampoule was then put inside of a single zone horizontal tube furnace with the FeSe powder staying at the center of the furnace. The furnace was then slowly heated up to 400$^\circ$C and kept at this temperature for two weeks. The temperature at the cold end is determined to be 375$^\circ$C. Figure\,\ref{FeSe}(b) shows a picture of the ampoule during growth. The residual starting materials at the hot end and a cluster of grown FeSe crystals at the cold end can be observed through the clear flux melt. After two weeks, the furnace was shut off and cooled to room temperature. Figure\,\ref{FeSe}(c) shows the picture of the ampoule after cooling to room temperature. We extended the growth to two months and found there is always some residual starting material at the hot end. The crystals were then washed out of flux and cleaned as described by Chareev et al.\cite{chareev2013single} Figure\,\ref{FeSe}(d) shows the cluster of crystals at the cold end and the picture of one piece of plate-like crystals on a mm grid. Our transport measurements under hydrostatic pressure confirmed that FeSe crystals grown in this way are of good quality.\cite{sun2016dome,sun2017high}

\subsubsection{van der Waals magnets and ferroelectrics}
\begin{figure} \centering \includegraphics [width = 0.47\textwidth] {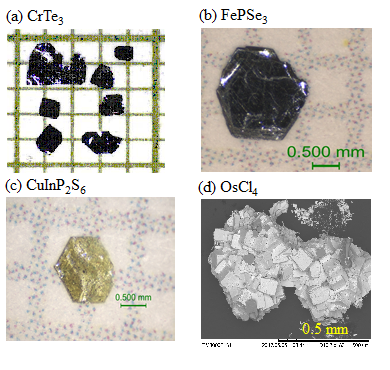}
\caption{(color online) Pictures of compounds grown out of a KCl-AlCl$_3$ eutectic flux: (a) CrTe$_3$ in a mm grid, (b) FePSe$_3$, (c) CuInP$_2$S$_6$, and (d) SEM picture of a cluster of OsCl$_4$ cubes deposited at the cold end.}
\label{MAS}
\end{figure}

The low melting temperature of KCl-AlCl$_3$ eutectic mixture and the successful growth of FeSe crystals suggest that KCl-AlCl$_3$ flux can be employed to grow other chalcogenides with low melting or decomposition temperatures. We thus tried the growth of CrTe$_3$ out of the KCl-AlCl$_3$ eutectic flux using the horizontal configuration. CrTe$_3$ is an antiferromagnetic semiconductor with magnetic layers made up of lozenge shaped Cr$_4$ tetramers. According to the Cr-Te phase diagram, CrTe$_3$ melts incongruently at 480$^\circ$C.  The peritectic plateau extends to 97\% Te. The liquid coexisting with CrTe$_3$ has a very narrow the composition range (97\%-98\% Te) in a very narrow temperature range (445-480$^\circ$C). This narrow composition and temperature ranges allow conventional vertical flux growth but with a rather low yield.\cite{mcguire2017antiferromagnetism} Liquid transport growth using self flux is expected to work well for this compound but with the concerns of low growth temperature and the difficulty of extracting the crystals from the residual Te flux. We thus tried the liquid transport growth out of the halide flux since CrTe$_3$ is not hygroscopic and crystals can be easily washed out of the halide flux.  In our growth, about 100\,mg of prefired CrTe$_3$ powder , KCl, and AlCl$_3$ were loaded into a silica tube (9 mm I.D., 1.5 mm wall thickness)  that was then sealed under vacuum. As for the growth of FeSe, the chemical loading was performed inside of a dry glove box filled with He. The ampoule was heated in a horizontal furnace with the hot end held at 450$^\circ$C and the cold end at 425$^\circ$C. Millimeter sized platelike crystals is obtained after several days to one week(see Fig.\,\ref{MAS} (a)). These CrTe$_3$ crystals have comparable in-plane dimension but are much thinner than those grown of the Te flux in the vertical configuration. The details of crystal growth and physical properties have been reported elsewhere.\cite{mcguire2017antiferromagnetism}

The same flux technique has been employed to grow other van der Waals functional compounds where two dimensional layers are held together by van der Waals bonding, such as transition metal selenophosphates (MPSe$_3$, M=transition metal) and thiophosphates (MPS$_3$).  In these
compounds the functionality is determined by the type of metal present on the cation sublattice, yielding materials with functionalities ranging from magnetic to ferroelectric. While the vapor transport method using I$_2$ as the transport agent can grow sizable single crystals of MPS$_3$, it does not work well for some MPSe$_3$ members as P-Se mixture melts at rather low temperatures. By using a similar procedure and the same temperature profile used for the growth of CrTe$_3$, we have been growing out of the KCl-AlCl$_3$ eutectic flux millimeter sized single crystals of NiPSe$_3$  and FePSe$_3$, and ferroelectric  CuInP$_2$S$_6$.  Figure\,\ref{MAS} (b) and (c) show the crystal pictures of FePSe$_3$ and CuInP$_2$S$_6$, respectively.

In the above growths of FeSe, CrTe$_3$, MPSe$_3$, and InCuP$_2$S$_6$, the KCl-AlCl$_3$ mixture does not change the composition of the starting materials and is expected to work as transport media only. In some special cases, the halide salt mixture can chemically react with the starting materials and deposit at the cold end small halide crystals. A good example of this is RuCl$_3$. $\alpha$-RuCl$_3$ has attracted intense attention recently as a candidate with proximate Kitaev quantum spin liquid behavior.\cite{banerjee2016proximate} Considering possible reaction between oxides and the halide flux, we tried the growth of RuCl$_3$ starting with RuO$_2$ powder in the KCl-AlCl$_3$ mixture. About 1 gram of RuO$_2$ was first dried at 900$^\circ$C for overnight and then sealed in a quartz tube together with the KCl-AlCl$_3$ eutectic mixture. The ampoule was heated in a horizontal furnace with the hot end at 420$^\circ$C and the cold end at 380$^\circ$C. After 8 days, plate-like crystals of submillimeter size form with the majority at the cold end and some in the middle of the ampoule. X-ray powder diffraction confirms that both $\alpha$- and $\beta$-RuCl$_3$ of comparable fraction are obtained.  We did not try to optimize the flux growth because large $\alpha$-RuCl$_3$ crystals up to 0.5 gram per piece can be obtained using the sublimation behavior of RuCl$_3$ at high temperatures.\cite{banerjee2016neutron} However, considering the commercial RuCl$_3$ powder is always contaminated with RuOCl$_2$, Ru, or RuO$_2$, reacting RuO$_2$ with the KCl-AlCl$_3$ eutectic flux can be employed to provide phase pure RuCl$_3$ starting materials for further crystal growth. We also tried to react OsO$_2$ with the KCl-AlCl$_3$ eutectic flux with the OsO$_2$ powder at the hot end of 400$^\circ$C. The temperature at the cold end is kept at 370$^\circ$C.  After 6 days, the flux turns to be brown with some black deposition at the cold end. After washing in water, only the black deposition is left. Figure\,\ref{MAS}(d) shows the SEM picture of a cluster of small cubes deposited at the cold end. X-ray powder diffraction at room temperature shows the black deposition is OsCl$_4$ (space group \textit{Cmmm}).\cite{cotton1977structure} As OsCl$_3$ is reported to be dark brown and hygroscopic,\cite{merten1962high} the color change of the flux suggests that OsCl$_3$ might form but is washed away. The liquid transport growth technique can be employed to grow halide crystals or to provide necessary starting materials for crystal growth. This approach can be considered when growing various layered transition metal halides.\cite{mcguire2017crystal}

In all the above growths using the KCl-AlCl$_3$ eutectic flux, the temperature of the hot end is kept around 400$^\circ$C. Higher growth temperatures above 500$^\circ$C might lead to tube failure possibly due to the reaction between quartz tube and halide flux and the vapor pressure of the halide flux.

\section{Discussion}

\begin{figure} \centering \includegraphics [width = 0.4\textwidth] {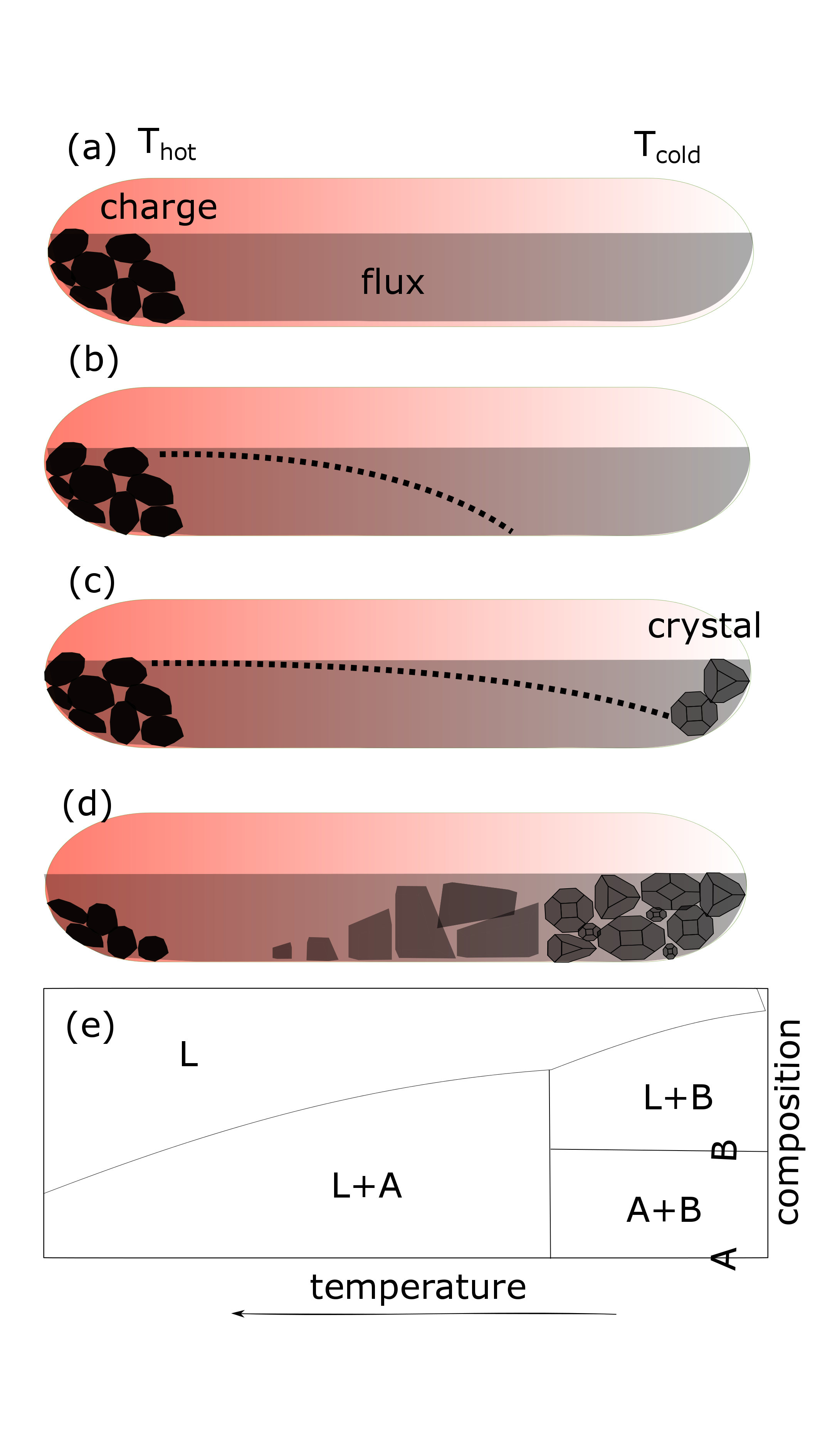}
\caption{(color online) Possible mechanism of liquid transport growth. (a) The growth ampoule when flux first melts with little charge dissolved. (b)The ampoule in early stage of the growth highlighting the diffusion of dissolved charge from the hot end to the cold end. The dashed curve shows possible composition gradient of charge in flux at a certain time. (c) Crystals start to precipitate at the cold end once the concentration of charge is above the solubility limit. Panel (d) and (e) illustrate that different crystals can be obtained by controlling the temperature profile.  (d) Different crystals (crystal A and B) can precipitate in one single growth depending on the temperature profile along the horizontal axis. (e) one hypothetical phase diagram that determines the stable phases at different positions in (d). }
\label{mechanism}
\end{figure}

\textit{Growth mechanism}
Figure\,\ref{Transport} illustrates the principles of vapor transport and liquid transport growth. The difference between these two techniques is the state of transport medium. For the vapor transport growth, a small amount of transport agent in a vapor state works as the transport agent at high temperatures. While for the liquid transport growth, the growth ampoule is partially filled with a large amount of flux and the flux melt is expected to transport the charge from the hot end to the cold end. For both techniques, crystals are expected to first appear at the cold end. This has been well established for vapor transport growth. Two observations in our proof of principle growths are important in understanding the growth mechanism of the liquid transport growth: (1) IrSb$_3$ (see Fig.\,\ref{IrSb}(c)), MnBi (Fig.\,\ref{MnBi}(a)), FeSe (Fig.\,\ref{FeSe}(c)), CuInP$_2$S$_6$, FePSe$_3$, and OsCl$_4$ single crystals are found at the cold end of the growth ampoule. This observation confirms that crystal growth first occurs at the cold end and suggests the flux melt behaves as the transport agent; (2) Mo$_3$Sb$_7$ crystals were found in the whole ampoule for the growth starting with a large amount of starting Mo. This indicates that the growth front gradually moves from the cold end to the hot end and crystallization can take place in the whole ampoule with enough growth time and starting materials.

With the above observation and analysis, the detailed growth process can be summarized as the schematic in Fig.\,\ref{mechanism}: (1) Fig.\,\ref{mechanism}(a) illustrates the state when the flux just melts with very little charge dissolved into the flux melt. (2) As time proceeds, the flux melt starts to dissolve the charge. The dissolved charge diffuses from the hot end to the cold end due to the compositional gradient. The dashed curve in Fig.\,\ref{mechanism}(b) shows the concentration of dissolved charge after some time of dwelling at high temperatures. (3) With more extended stay at high temperatures, the concentration of the dissolved charge at the cold end goes beyond the solubility limit. Under enough supersaturation, nucleation occurs and the nuclei grow with the charge transported from the hot end by the flux melt as illustrated in Fig.\,\ref{mechanism}(c). With the grown crystals taking the space at the cold end, the growth fronts gradually moves to the hot end.

\textit{Temperature is critical}
Temperature is one critical parameter in liquid transport growth as in other high temperature growths. Thermodynamically, it determines the phase that will precipitate out of the melt. In the growth of Mo$_3$Sb$_7$ which is the only compound in the Mo-Sb binary system, Mo$_3$Sb$_7$ is expected to nucleate as long as the temperature at the cold end is above the melting temperature of Sb. However, for systems with more than one compounds, such as Ir-Sb (Fig.\,\ref{IrSb}) or Mn-Bi (Fig.\,\ref{MnBi}), the growth temperature should be carefully selected to crystallize the desired phase. By controlling the temperature gradient along the horizontal axis, it is possible to crystallize different phases at different positions of the growth ampoule. This is well illustrated with Fig.\,\ref{mechanism}(d) and (e). Fig.\,\ref{mechanism}(e) shows a hypothetical phase diagram with two phases A and B which would grow out of flux melt in different temperature intervals. Fig.\,\ref{mechanism}(d) illustrates that crystals A and B can grow above and below the peritectic temperature, respectively, in one single liquid transport growth.

Kinetically, the temperature determines the solubility and diffusivity of solute inside of the melt and the supersaturation, thus the morphology and dimension of the crystals. The growth process illustrated in Fig.\,\ref{mechanism} suggest that the diffusion transport of solute in the flux melt is prerequisite for the crystallization. We can thus expect that the growth rate is proportional to the diffusion coefficient of solute in the flux melt, and the solubility difference between the hot end and the cold end, and inversely proportional to the diffuse length between the charge and the growth front. The values of solubility for given temperatures can be taken from the liquidus curve. From the phase diagrams shown in Fig. \ref{IrSbPD}, \ref{MoSb},  and \ref{MnBiPD}, the compositional gradient for the growth of MnBi is much larger than that for Mo$_3$Sb$_7$ and IrSb$_3$. The fact that MnBi crystals are small suggests that the diffusion coefficient, which is a function of both temperature and position in liquid transport growth, is the major factor affecting this growth. The small crystals and quite some Bi inclusions might both result from the low growth temperature below 300$^\circ$C. At such a low temperature, the viscosity of Bi is large and the diffusivity of solute is small. This is supported by the observation in a conventional flux growth that stirring the melt by shaking the crucible whelps to produce larger crystals.\cite{williams2016extended}

\textit{Temperature gradient is necessary}
In all our liquid transport growths, a temperature gradient along the horizontal axis is always applied and this temperature gradient is believed to be necessary for the completion of the growth. Without the temperature gradient, a homogeneous melt is expected at high temperatures and random nucleation would occur once the charge concentration is beyond the solubility limit. The temperature gradient along the horizontal axis makes it possible for the nucleation first occurs at the cold end and the growth front gradually moves from the cold end to the hot end of the growth ampoule. The observation of IrSb$_3$ (Fig.\,\ref{IrSb}) and MnBi (Fig.\,\ref{MnBi}) crystals only in the 2-3\,cm range from the cold end supports that the crystallization first occurs at the cold end and then gradually moves to the hot end.

However, the temperature gradient might affect the stoichiometry of the as-grown crystals for those compounds with a temperature dependent composition. For example, the fraction of Te in 1T'-MoTe$_2$ changes from 66\%at to 68\%at when cooling from 1180$^\circ$C to 998$^\circ$C according to the available phase diagram. Liquid transport growth with horizontal temperature gradient is not ideal for 1T'-MoTe$_2$ as the growth will result in crystals with different stoichiometry and thus different physical properties. This is supported by the fact that MoTe$_2$ crystals grown at a fixed temperature show a larger residual resistivity ratio and magnetoresistance than those grown by cooling the self flux in the temperature range 998$^\circ$C$\sim$1180$^\circ$C.\cite{huang2016spectroscopic} The temperature gradient may also affect the composition of dopants if the distribution coefficient of the dopant has a large temperature and composition dependence. This effect should be taken into consideration when liquid transport technique is employed to grow some doped compositions. On the other hand, this effect of the temperature gradient can be employed to fine tune the stoichiometry of crystals.

For all the growths reported in this work, the charge always stays at the hot end. However, if the solubility increases with decreasing temperature, the charge should be kept at the cold end. This is analogous to what happens in vapor transport growth: the starting materials are kept at the cold end and crystals form at the hot end of the growth ampoule when the reaction between the charge and the transport agent is exothermic; the temperature gradient should be reversed if the reaction is endothermic.

\textit{Comparison with the vertical flux growth}
Compared to the conventional vertical flux growth, the following features of the liquid transport growth are noteworthy:

(1) The charge/flux ratio is not limited by the solubility limit at given temperatures. In the vertical flux growth, the charge/flux ratio is determined by the liquidus line at given temperatures.  In contrast, a much larger quantity of charge can be used in the liquid transport growth. This feature allows the liquid transport growth to produce a large quantity of crystals in a single growth. For all three incongruent melting compounds grown out of a self flux reported in Section III A, the composition of the peritectic liquid coexisting the desired phase gives the maximum fraction of charge allowed without precipitating other phases in a vertical flux growth. This can limit the yield of crystals in a single growth as described below.

(2) Crystal growth occurs before fully dissolving the charge. For the vertical flux growth, it is important to stay at temperatures above the liquidus line at a given composition to form a homogeneous melt before cooling. For the liquid transport growth, the flux dissolves the charge and the compositional gradient along the horizontal axis drives the diffusion transport.

(3) A temperature gradient is applied along the horizontal axis and mixing of charge and flux should be avoided. This temperature gradient is important in driving the diffusion transport of charge from the hot end to the cold end of growth ampoule and driving the crystal growth in opposite direction. Random or uniform distribution of charge in flux could impede the diffusion transport of charge and might induce random nucleation inside of the growth ampoule.

It should be noted that most compounds mentioned in this work can be grown out of flux using the conventional vertical configuration. The conventional vertical flux growth is still the more widely used technique which should be considered first. The liquid transport growth is a complementary technique and is recommended when a large amount of crystals are needed but the yield from the conventional flux growth is limited. Both IrSb$_3$ and Mo$_3$Sb$_7$ melt incongruently and the peritectic plateau extends to very high Sb content according to the available phase diagrams. This limits the fraction of charge in the melt thus the yield of crystals when the conventional vertical flux is performed. For example, a Mo:Sb ratio of 1:49 is used for the growth of Mo$_3$Sb$_7$ and a Ir:Sb ratio of 3:97 for IrSb$_3$. MnBi is a good example that the low yield from the vertical flux growth is caused by the large compositional difference between the solid phase and the coexisting liquid phase. According to the Mn-Bi phase digram, the composition of MnBi (50\% Mn) is quite different from that of the coexisting liquid (3-10\% Mn). Following the lever rule, this limits the yield of a conventional vertical flux growth.  There is no phase diagram or other thermodynamic data about the solubility of chalcogenides or halides in the KCl-AlCl$_3$ eutectic flux. However, the observation that there is always some leftover starting materials in the hot end suggests that the solubility is limited. Liquid transport growth of chalcogenides and halides  in the KCl-AlCl$_3$ eutectic flux can produce a large amount of crystals in a single growth as it is not limited by the solubility at given temperatures.

\textit{Extracting crystals from the flux}
For liquid transport growth from a halide flux, the flux can be washed away in water or other solvents in case the crystals are moisture sensitive. For liquid transport growth from a metal flux, we tried two different ways to separate the crystals from the flux. For IrSb$_3$, we loaded the part from the cold end of the liquid transport growth ampoule into a Canfield crucible set, and centrifuged it after melting the flux at high temperatures. This approach does not work if more than one phase is obtained (Fig.\,\ref{mechanism}(d)) in the ampoule or the stoichiometry of crystals is sensitive to the temperature gradient. For Mo$_3$Sb$_7$, the flux wrapping the crystals can be removed mechanically using a surgical blade since the crystals are large and mechanically strong. This mechanical extraction does not work for soft crystals such as MoTe$_2$, which is soft and exfoliable. Other methods of extracting crystals from solid flux might be needed.\cite{wolf2012flux}

\emph{This is not horizontal Bridgman}
This growth technique described in this work is different from the horizontal Bridgman method, or modified Bridgman technique performed in a horizontal temperature gradient, or the horizontal gradient-freeze technique\cite{schunemann2006horizontal}. The fundamental difference lies in whether the charge and flux are well mixed and a homogeneous melt is obtained before generating supersaturation. As in the vertical flux growth, the Bridgman technique involves melting all starting materials (both charge and flux for the modified Bridgman technique) at high temperatures to form a homogeneous melt before applying a temperature gradient to generate supersaturation. For the liquid transport growth, the amount of charge is always more than the flux can dissolve at the growth temperature and a homogeneous melt is never obtained before precipitation occurs at the cold end.

The other prominent difference is the presence of a compositional gradient along the horizontal axis. For the modified Bridgman technique performed in a horizontal configuration, the melt far away from the growth front is homogeneous in composition and the compositional gradient exists only near the growth front. The nucleation and growth takes place once the temperature of the cold end is below the liquidus line. In contrast, the compositional gradient in the horizontal transport technique is inherent and the crystallization takes place once the concentration of solute is above the solubility limit. The magnitude of the composition difference along the horizontal axis depends on the slope of the liquidus line in phase diagram.

\textit{Possible role of vapor phase}
It is worth mentioning that the sealed ampoules are only partially filled with the flux. It is likely the vapor phase might also play a role in species transport. In the growth of IrSb$_3$, Mo$_3$Sb$_7$, and MnBi, the vapor pressure of starting materials or other intermediate binary phases are small at the growth temperatures employed in the growth. Single crystals are found inside of the flux. For the growth out of halide flux, according to the pseudo-binary AlCl$_3$-KCl phase diagram, there is a certain vapor pressure around 400$^\circ$C. This vapor pressure might be involved in the mass transport in the growth of FeSe as FeSe crystals are found both below and above the melt at the cold end. Because of this, Bohmer \textit{et al.} called growth using this technique as vapor transport growth and they purposely tilted the growth ampoule.\cite{bohmer2016variation} However, in all other growths we performed, single crystals are observed only inside of the melt which does not suggest a significant role of vapor phases in the growth. The possible effect or involvement of the vapor phases deserves further investigation in the liquid transport growth.

\section{Summary}

In summary, we report the flux growth in a horizontal configuration with a temperature gradient along the horizontal axis. At high temperatures, the flux melt dissolves the charge at the hot end and transports to the cold end where crystallization occurs after the concentration of solute is above the solubility limit. The growth principles involved are similar to those in the vapor transport growth. We thus name it liquid transport growth. This technique can provide a large amount of crystals in a single growth as the charge/flux ratio is not limited by the solubility limit at given temperatures.  This is supported by the growths of proof-of-principle compounds reported in this work. This technique can grow crystals of comparable dimension as the conventional flux growth. Temperature is a key growth parameter which determines the phases that would precipitate and the crystal size. The temperature gradient along the horizontal axis is needed to drive the diffusion transport from the hot end to the cold end of the growth ampoule and to drive the crystal growth in an opposite direction.

Our proof-of-principle compounds reported in this work either melt incongruently or have a large vapor pressure at high temperatures. We would expect the liquid transport growth can also be employed to grow congruent melting compositions at easily accessible temperatures much lower than the melting point. We also expect that this technique works well for the flux growth of oxides. This liquid transport technique is complementary to the conventional vertical flux growth and can be employed when a large amount of crystals are needed but the yield from the conventional vertical flux growth is limited. When the temperature has to be low in order to precipitate the right phase, the large viscosity of the melt and the low diffusivity of solute in the melt might limit the size of the final crystals. This concern should be taken into consideration when selecting the appropriate configuration, horizontal vs vertical, for the flux growth of desired crystals.

The present work is more focused on understanding the principles of liquid transport growth. We did not take efforts to optimize the growth conditions. More future studies are needed to understand the effects of some important growth parameters, such as temperature and temperature gradient along the horizontal axis, the flux filling of the growth ampoule, the dimension of the ampoule, the charge/flux ratio, and the tilting angle of the growth ampoule. Our present understanding of liquid transport growth suggests that the diffusion of the solute in the flux melt might control the growth rate. However, dissipation of the latent heat away from the crystal surface or reactions at the crystal-melt interface could significantly affect the growth rate in some cases. The kinetics of liquid transport growth also deserves further study.

\section{Acknowledgment}
This research was supported by the U.S. Department of Energy, Office of Science, Basic Energy Sciences, Materials Sciences and Engineering Division. Manuscript preparation was partially funded by the Air Force Research Laboratory under an Air Force Office of Scientific Research grant (LRIR No. 14RQ08COR) and a grant from the National Research Council.

This manuscript has been authored by UT-Battelle, LLC, under Contract No.
DE-AC0500OR22725 with the U.S. Department of Energy. The United States
Government retains and the publisher, by accepting the article for publication,
acknowledges that the United States Government retains a non-exclusive, paid-up,
irrevocable, world-wide license to publish or reproduce the published form of this
manuscript, or allow others to do so, for the United States Government purposes.
The Department of Energy will provide public access to these results of federally
sponsored research in accordance with the DOE Public Access Plan (http://energy.gov/
downloads/doe-public-access-plan).



%

\end{document}